\documentstyle[aps,pra,twocolumn,floats,psfig]{revtex}

\newcommand{\Tr}{{\hbox{Tr}}}

\begin{document}

\draft
\title{An open systems approach to calculating time dependent spectra}
\author{Martti Havukainen and Stig Stenholm$^{\dag}$
\footnotetext{$^{\dag}$Permanent affiliation: Physics Department, Royal Institute of Technology, Stockholm Sweden}}
\address{Helsinki Institute of Physics, P. O. Box 9, FIN-00014 University of Helsinki, Finland}
\date{\today}
\maketitle

\begin{abstract}
A new method to calculate the spectrum using cascaded open systems and master equations is presented.
The method uses two state analyzer atoms which are coupled to the system of interest, whose spectrum
of radiation is read from the excitation of these analyzer atoms. The ordinary definitions of a spectrum
uses two-time averages and Fourier-transforms. The present method uses only one-time averages.
The method can be used to calculate time dependent as well as stationary spectra.
\end{abstract}


%\twocolumn

\section{INTRODUCTION}
\label{intro}
%\narrowtext
Much of the information obtained in laser experiments comes in the form of spectral data. In steady
state, these are supposed to give the Fourier transform of the energy level structure of the system
under investigation. In this manner we have been able to learn about the quantum configurations of
atoms, molecules and solid state systems.

Today, however, much work is done with pulsed laser sources, where the pulse duration samples the
evolution of the system over times ranging from nanoseconds to femtoseconds. Spectral data are still
recorded, but their significance and physical information is no longer straightforward. Thus one needs
to reconsider the definition and computation with due consideration of the physical conditions under
which the data are obtained. Several attempts have been made to modify the Fourier-transform
definition, the Wiener-Khintchine spectrum \cite{page,lampard,silverman}, but here we choose to
consider the physical spectrum introduced by Eberly and W\'odkiewicz \cite{eberly}, which attempts to
emulate the spectral measurements using e.g a Fabry-Perot filter in front of the detector. They have
applied it to the well known case of the fluorescence Mollow spectrum \cite{eberlymollow}. A very
similar approach has allowed Kowalczyk et al. \cite{Kowalczyk} to consider the fluorescent spectra
deriving from a laser-excited molecular wave packet. A model calculation of such a process has been
presented by Vinogradov and Janszky \cite{Vinogradov}. Another approach to the molecular situation is
presented by Lee et al. \cite{Lee}.

In this paper we consider the problem of obtaining spectral data in an evolving system. Thus only the
accumulated information is available; the future is still unknown. This rules out the use of a full
Fourier-transform, and only physically manipulated collected data can be used. The spectral
measurements require a filtering, which smeares the signal in time. The frequency-time resolution
has to obey an uncertainty relation. In addition, the exact nature of the transfer of spectral
information from the system investigated to the detector imposes its own limitations. There can be
no unique spectral definition for time-dependent systems, but we can require that all definitions agree
when infinite measurement times are available.

In a laboratory experiment, the radiation emitted from a driven system reaches the detector through
a technical setup which, in addition to its function as a filter, will impose its own noise limits.
The filtering action provides a noisy channel. This is, however, just the situation which is described by
the term ``Open system'' \cite{carmichael}. We want to consider an open systems approach to
time-dependent spectra.

A natural frequency-selective detector is a two-level atom. This will respond to radiation only within
a bandwidth given by its natural linewidth. If this is small enough, the spectrum if incoming radiation
is resolved with this accuracy; this mirrors the quantum fluctuations of the detector-atom decay. We
also let the radiation reach the detector atoms through a noisy channel, a reservoir which can be
eliminated. Thus we present a new approach to time dependent spectra, which does not seem to be directly
related to the ones used earlier. The aim of the present paper is to introduce this model and compare
it with the physical spectrum of Eberly and W\'odkiewicz and the Wiener-Khintchine spectrum for steady
state.

In Sec. II.A we introduce the system we want to investigate the spectrum of.
We choose a simple one consisting
of three levels only. In fact, we would like to consider some real situation like an excited molecule but the
computational burden imposed by the theoretical methods is so large, that we have been forced to carry out our
comparisons on this simple model.

Sec. II.B.1 explains how the stationary Wiener- Khintchine spectrum can be calculated in the Schr\"odinger
picture using the quantum regression theorem, this is used as a reference in the following. In Sec.II.C.1 it is
shown how the quantum regression theorem can be used to obtain the physical spectrum of Eberly and W\'odkiewicz.
We will use these results to judge the reliability of the method presented in this paper. Sec. II.C.2 presents
the theory of our own approach, which relies heavily on the work by Carmichael \cite{carmichael}. The derivation
leads to a master equation, which is used in a quantum simulation to obtain spectra as explained above. The
results are presented in Sec. III and compared with the physical spectra in Sec. III.C. Finally Sec. 
\ref{conclusion} contains the conclusions.

\section{MEASURING THE TIME-DEPENDENT SPECTRUM}

\subsection{ The system under investigation}
\label{systemunderinvestigation}
The model system we choose for our investigation is the 3-level atom shown in Fig. 1. The three levels are taken
to have the (dimensionless) energies: $\omega_1=0$, $\omega_2=4.0$ and $\omega_3=8.0$. The levels $|2\rangle$
and $|3\rangle$ decay to level $|1\rangle$ by the rates $\Gamma_1=\Gamma_2=0.1$. 
These parameters are kept constant all through the present paper.
The level pairs $|2\rangle <-> |1\rangle$ and $|3\rangle <-> |2\rangle$ are coupled through lasers with the Rabi
frequencies $\Omega_1$ and $\Omega_2$ respectevely. These coupling strengths are varied in the calculations. The
fact, that the dipole approximation does not allow all transitions considered, lacks
significance for the features we are investigating in this paper. In a molecular system parity may not be a good
quantum number, in an atom or quantum dot system, the decay $\Gamma_2$ may derive from higher multipole
transitions.

\begin{figure}[htp]
\vspace{-2cm}
\centerline{\psfig{file=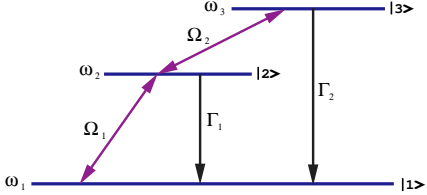,width=8.5cm,bbllx=1cm,bblly=1cm,bburx=4cm,bbury=4cm,clip=}}
\vspace{-2.5cm}
\caption{The three level atom used in the calculations. The atom has three energy levels
$\omega_1=0.0$, $\omega_2=4.0$ and $\omega_3=8.0$, two internal decays $\Gamma_1=\Gamma_2=0.1$ and
two lasers $\Omega_1$ and $\Omega_2$.}
\label{threelevel}
\end{figure}

Mathematically the dynamics of this kind of system can be studied using master equations. We have included
two different spontaneous decays to the reservoirs so the master equation has two decay terms of the Lindblad form.
The reservoir is taken to be at zero temperature. Assuming the Born-Markov and the Rotating Wave Approximations 
(RWA) we find the master equation

\begin{equation}
\label{mastereq}
\frac{d\hat{\varrho}}{dt}=-i[\hat{H},\hat{\varrho}] + \sum\limits_{i=1}^2
\Gamma_i(\hat{L}_i\hat{\varrho}\hat{L}_i^{\dag} -
\frac{1}{2}\hat{L}_i^{\dag}\hat{L}_i\hat{\varrho} - \frac{1}{2}\hat{\varrho}\hat{L}_i^{\dag}\hat{L}_i)
% +
% \frac{\Gamma_1}{2}(\hat{L}_2\hat{\varrho}\hat{L}_2^{\dag} + 
% \frac{1}{2}\hat{L}_2^{\dag}\hat{L}_2\hat{\varrho} + \frac{1}{2}\hat{\varrho}\hat{L}_2^{\dag}\hat{L}_2
\end{equation}
where $\hat{L}_1=|1\rangle\langle 2|$ and $\hat{L}_2=|1\rangle\langle 3|$ are Lindblad operators. Everywhere
in this paper the units have been chosen in such a way that $\hbar =1$. Lasers are taken into account by
adding resonant driving terms to the Hamiltonian; the laser frequencies are chosen such that

\begin{eqnarray}
\label{laserfreqs}
\omega_{l1} & = & \omega_2-\omega_1, \nonumber\\
\omega_{l2} & = & \omega_3-\omega_2.
\end{eqnarray}

In the rotating-wave approximation the Hamiltonian becomes

\begin{eqnarray}
\label{hamilton0}
\hat{H}=\sum\limits_{i=1}^{3}\omega_i |i\rangle\langle i| + \frac{\Omega_1}{2}(|1\rangle\langle 2|
e^{i\omega_{l1}t} + {\rm h.c}) + \nonumber\\
\frac{\Omega_2}{2}(|2\rangle\langle 3|e^{i\omega_{l2}t} + {\rm h.c}).
\end{eqnarray}

\subsection{The measurement scheme and the stationary spectrum}
\begin{figure}[htp]
\vspace{-2cm}
%\centerline{\psfig{file=kuva3level4.epsi,width=8.5cm,bbllx=0cm,bblly=0cm,bburx=25cm,bbury=25cm,clip=}}
\centerline{\psfig{file=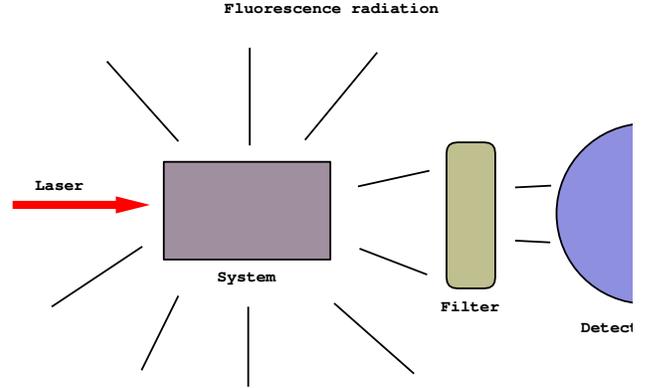,width=8.5cm,bbllx=0cm,bblly=0cm,bburx=25cm,bbury=25cm,clip=}}
\vspace{0.2cm}
\caption{The measurement scheme. The system is driven by a laser. Part of the fluorescence
radiation passes the filter to the detector.}
\label{filterfigure}
\end{figure}

The usual setup for spectrum measurements is presented in Fig. \ref{filterfigure}. Fluorescence radiation from
the three level atom is guided to the filter which allows radiation with only a certain frequency
$\omega_D$ to go through it. Behind the filter, a photodetector detects the intensity of the
radiation coming through. The spectrum can be measured by scanning $\omega_D$ over an appropriate
frequency range and recording the relative intensities. The same measurement configuration can be used to measure
both stationary and time-dependent spectra.

\subsubsection{Stationary spectrum and the Quantum Regression Theorem}
\label{wkqrt}
Mathematically the spectrum for a stationary system, the Wiener-Khintchine spectrum, is defined as

\begin{equation}
\label{wk}
S_{WK}(\omega)=\int\limits_{-\infty}^{\infty}d\tau\langle\hat{a}^{\dag}(t)\hat{a}(t+\tau)\rangle
e^{i\omega\tau},
\end{equation}
where $\hat{a}$ and $\hat{a}^{\dag}$ are field operators. The definition contains two-time averages of the field
modes. These expectation values are averages over the degrees of freedom of the system, so in practise the
spectrum is calculated from two-time system expectation values. 
% EHKA TARKEMPI SELITYS

In the calculation of the spectrum, two-time averages of system operators
in the Heisenberg picture are needed. Simulations are, however, done in the Schr\"odinger picture. The method which
allows us to calculate multitime averages in the Heisenberg picture using a master equation is called the
Quantum Regression Theorem (QRT).

The calculation of the expectation value
$\langle \hat{O}_2(t+\tau)\hat{O}_1(t)\rangle,\ \tau\ge 0$ can be carried out with the QRT using the following steps:

1. Evolve the density matrix $\hat{\varrho}$ using the master equation from the initial time $t_0$ 
to the time $t$.

2. Form the matrix $C_{ij}(t,\tau=0)$

\begin{equation}
\label{c}
C_{ij}(t,\tau=0)=\Tr[\varrho (t)|i\rangle\langle j|\hat{O}_1],
\end{equation}
where $\hat{O}_1$ is the operator in the Schr\"odinger picture.

3. Evolve $C_{ij}(t,\tau)$ using equation (\ref{mastereq}) with the initial matrix $C_{ij}(t,\tau=0)$

\begin{equation}
\label{equationc}
\frac{dC_{ij}(t,\tau)}{d\tau}=G_{ijkl}C_{kl}(t,\tau);
\end{equation}
here repeated indices are summed over.
Equation (\ref{equationc}) is the master equation (\ref{mastereq}) in component form. The coefficients $G_{ijkl}$ are
thus determined by the master equation.

4. Form $\langle\hat{O}_2(t+\tau)\hat{O}_1(t)\rangle$ using the matrix elements of $\hat{O}_2$ in the Schr\"odinger
picture ie.

\begin{equation}
\label{O2}
\langle\hat{O}_2(t+\tau)\hat{O}_1(t)\rangle=\hat{O}_{2,ij}C_{ij}(t,\tau).
\end{equation}
This procedure gives the expectation values for such time pairs $(t,t+\tau)$ where the operator $\hat{O}_2$ has 
a larger time value than the operator $\hat{O}_1$, ie. $\tau\ge 0$. In order to get time values in the reverse
order, we have to take a complex conjugate.
It is also possible to calculate expectation values of the type
$\langle\hat{O}_1(t)\hat{O}_2(t+\tau)\hat{O}_3(t)\rangle$. The difference is that at step 2 we have
 to calculate

\begin{equation}
\label{cc}
C_{ij}(t,\tau=0)=Tr[|i\rangle\langle j|\hat{O}_1\varrho (t)\hat{O}_3]
\end{equation}
and then continue as in the previous case.

\subsubsection{The Stationary spectrum for the three level system}

Without performing any calculations, we expect the following kind of structure of the stationary spectrum of
spontaneous emission for the three level system. When the laser
amplitudes are small, the spectrum should have two peaks whose relative intensities depend on the parameters
$\Gamma_1$ and $\Gamma_2$. If $\Gamma_1$ is large, then only level two should decay significantly and
the peak from transition from level three to one should be small. When the laser intensities are
increased, we should see some kind on Rabi splitting in the spectrum, ie. the two peaks are expected to display
substructure.

\begin{figure}[htp]
\vspace{0cm}
\centerline{\psfig{file=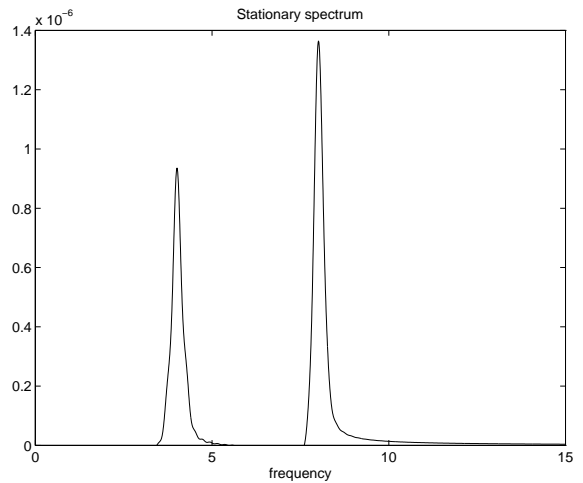,width=8.5cm,bbllx=1cm,bblly=1cm,bburx=20cm,bbury=23cm,clip=}}
\vspace{-2.4cm}
\caption{The stationary spectrum for our three level atom when lasers are weak. The energy levels and decay
constants are the same as in Fig. \protect\ref{threelevel}. The laser amplitudes are $\Omega_1=\Omega_2=0.2$.}
\label{stat2peak}
\end{figure}

\begin{figure}[htp]
\vspace{-0.2cm}
\centerline{\psfig{file=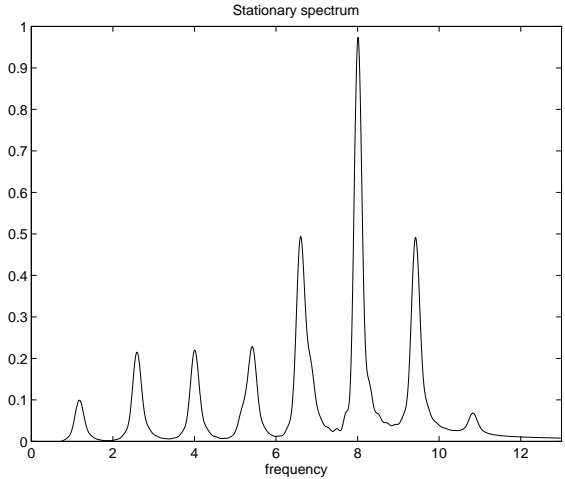,width=8.5cm,bbllx=1cm,bblly=1cm,bburx=20cm,bbury=23cm,clip=}}
\vspace{-2cm}
\caption{The stationary spectrum for our three level atom when the lasers are strong. The energy levels
and decay constants are the same as in Fig. \protect\ref{threelevel}.
The laser amplitudes are $\Omega_1=\Omega_2=2.0$.}
\label{stat8peak}
\end{figure}

The detection operator used in the calculations is

\begin{equation}
\label{operator1}
\hat{O}=\Gamma_1|1\rangle\langle 2| + \Gamma_2|1\rangle\langle 3|.
\end{equation}
The two-time average $\langle\hat{O}^{\dag}(t)\hat{O}(t+\tau)\rangle$ has been calculated and the spectrum
is obtained by integrating this expectation values with respect to $\tau$.
In Figs. \ref{stat2peak} and \ref{stat8peak} the stationary spectrum of our three level system is shown
for two different laser
amplitudes. In Fig \ref{stat2peak} the lasers are weak and we see two narrow peaks from the two transitions.
In Fig. \ref{stat8peak}
the lasers are ten times stronger compared to the first case. The energy levels are split and the spectrum
has got an eight peak structure, which can be understood as follows:
Because of the strong lasers all three energy levels are
split into sublevels. At low intensity we have observed two spectral components, at higher driving fields each
one is seen to split into four subcomponents. The frequencies are different for all four transitions.
This corresponds exactly to the number of degrees of freedom in a normalized 3$\times$3 density matrix.
The two time averages have an oscillating structure when the time separation $\tau$ is large. In
practical calculations, infinities in equation (\ref{wk}) have been replaced by large time values. As a result
we get asymmetries in the spectra and also finite linewidths.

\subsection{The physical spectrum}

One weakness of the WK-spectrum is that it is only defined if the system is in a stationary state.
This is clearly a limitation. For example, if instead of having a driving laser with constant intensity,
we took
a laser whose intensity changes in time, we obtain fluorescence light with varying intensity and
spectrum. The fluorescence light displays a spectrum but we cannot calculate it using the WK-definition.
There are several suggested definitions of a spectrum which would be physical also for nonstationary systems 
\cite{page,lampard,silverman}.
In the stationary limit all these spectra give the WK-spectrum.

A generalization which takes into
account also the measurement scheme was proposed by Eberly and W\`odkiewicz \cite{eberly}.
What they call the physical spectrum, is defined as

\begin{eqnarray}
\label{physspec}
\lefteqn{S_{PHYS}(t,\omega _f,\Gamma _f)=\int\limits_{-\infty}^{\infty}dt_1
 \int\limits_{-\infty}^{\infty}dt_2 H^*(t-t_1,\omega_f,\Gamma_f)\cdot} \nonumber\\
& & \hspace{2cm}H(t-t_2,\omega_f,\Gamma_f) \langle V^*(t_1)V(t_2)\rangle,
\end{eqnarray}
where $H(t,\omega_f,\Gamma_f)$ is a filter function characteristic of the filter (see Fig.\ref{filterfigure})

\begin{equation}
\label{filterfourier}
H(t,\omega_f,\Gamma_f)=\int\limits_{-\infty}^{\infty}d\omega H(w,\omega_f,\Gamma_f)
e^{-i\omega t}d\omega.
\end{equation}
$H(\omega,\omega_f,\Gamma_f)$ is a function of $\omega$, the parameter $\omega_f$ is the mean value of the
passband of the filter and $\Gamma_f$ is its width.
The filter function of a Fabry-Perot filter, which is used in our calculations, has the form

\begin{equation}
\label{fabryperot}
H(t,\omega_f,\Gamma_f)=\Theta (t)\Gamma_f \exp(-(\Gamma_f +i\omega_f) t),
\end{equation}
which gives for physical spectrum the expression

\begin{eqnarray}
\label{physspec2}
\lefteqn{S_{PHYS}(t,\omega_f,\Gamma_f) = \Gamma_f^2\int\limits_{-\infty}^{t}dt_1
 \int\limits_{-\infty}^{t}dt_2} \nonumber\\ &&\hspace{0.5cm} e^{-(\Gamma_f -i\omega_f)(t-t_1)}
e^{-(\Gamma_f +i\omega_f)(t-t_2)} \langle V^*(t_1)V(t_2)\rangle= \nonumber\\
&&\hspace{0.5cm} \Gamma_f^2\int\limits_{0}^{\infty}d\tau_1
 \int\limits_{0}^{\infty}d\tau_2 e^{-(\Gamma_f-i\omega_f)\tau_1}e^{-(\Gamma_f+i\omega_f)\tau_2}\cdot \nonumber\\
&& \hspace{3.5cm}\langle V^*(t-\tau_1)V(t-\tau_2)\rangle .
\end{eqnarray}

From the definitions above, we see that as well as for the Wiener-Khintchine spectrum, two time averages
are needed. The difference compared with $S_{WK}$ is that now we need only correlation functions whose times are
less than the time at which we calculate the spectrum.

\subsubsection{Calculation of the physical spectrum using QRT}

In order to calculate the physical spectrum at time $T$, we need to calculate the correlation functions
$\langle\hat{O}^{\dag}(t_1)\hat{O}(t_2)\rangle,\ \  0\leq t_1,t_2\leq T$. 
For numerical calculations we have
to discretize the time $t_1,t_2=n\cdot\Delta t,\ \ n=0...N,\ \ \Delta t=\frac{T}{N}$. There are now $N$ time
instants in between $t=0$ and $t=T$. The algorithm is
the following:

1.At $t=0$, form $C_{ij}(t_2=0,t_1=0)$ using the initial density matrix $\varrho (0)$ and use the algorithm
 presented in the last section to calculate $C_{ij}(t_2,t_1)$ for the time values 
$t_2=0,\ \ 0\leq t_1\leq T$.

2. Evolve $\hat{\varrho}(t)$ to the time $\Delta t$, form $C_{ij}(\Delta t,t_1)$ and use QRT to
evolve it to $t_1=T$. This gives the correlations at $t_2=\Delta t,\ \ \Delta t\leq t_1\leq T$.

3. Evolve $\varrho (t)$ to the next time step, form $C_{ij}$ with a new $t_2$ and evolve it to $t_1=T$.
Repeat this
step until $t_2=T$.

4. The algorithm above gives correlations for time pairs 
$t_2=n\cdot\Delta t, \ t_2\leq t_1\leq T,$  ie.  $t_2\leq t_1$. In order to obtain results for time values in
the reverse order, take the complex conjugate $C(t_1,t_2)^*=C(t_2,t_1)$.

Once we have calculated the correlation functions on a grid dense enough, the physical spectrum is a sum over
the grid according to equation (\ref{physspec}).

\begin{figure}[htp]
\vspace{-0.5cm}
\centerline{\psfig{file=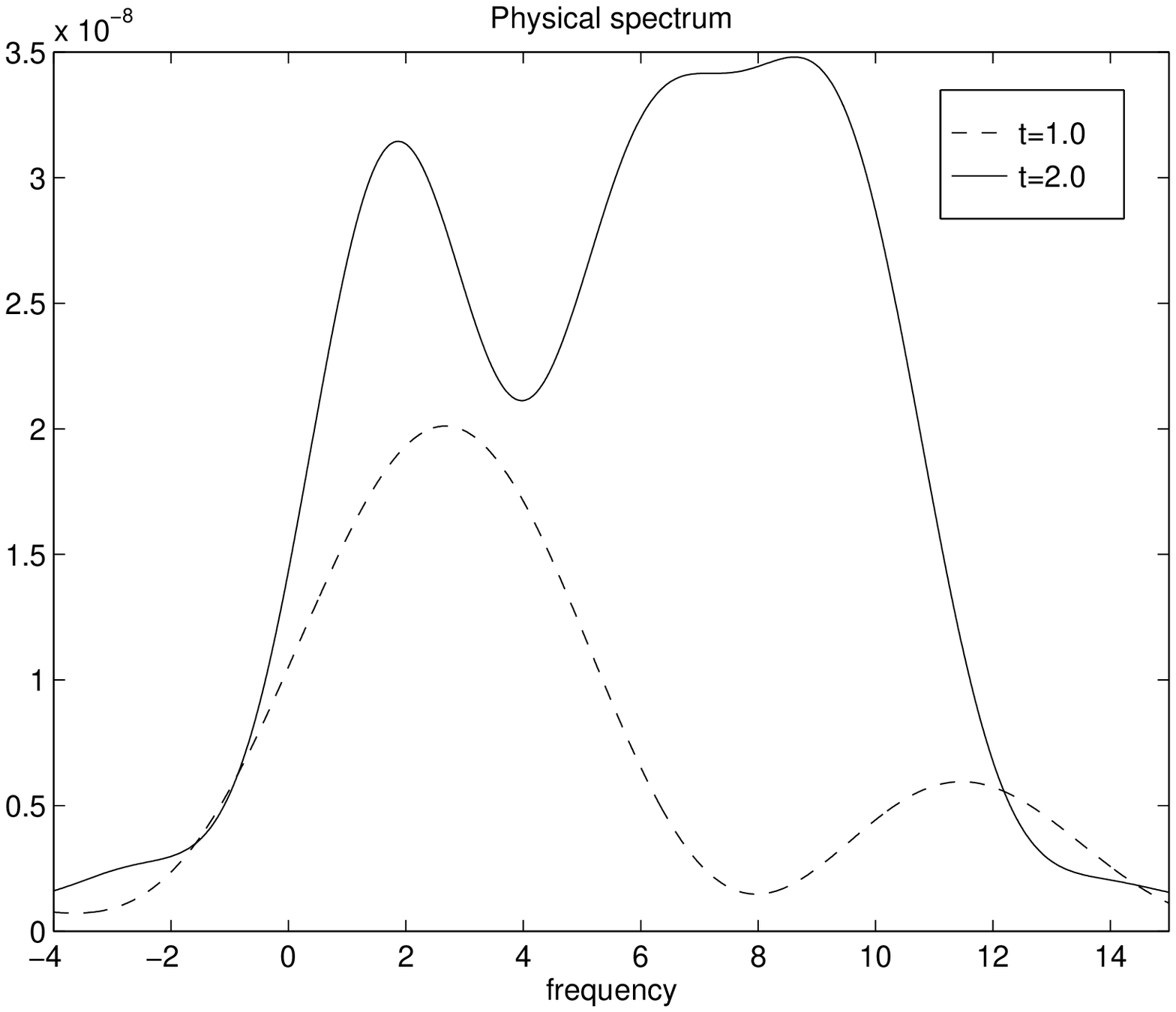,width=8.5cm,bbllx=1cm,bblly=1cm,bburx=20cm,bbury=23cm,clip=}}
\vspace{-2.3cm}
\caption{The time dependent physical spectrum calculated using QRT at the times $t=1.0$ and $t=2.0$.
The parameters as in Fig. \protect\ref{stat8peak}.}
\label{physspec1and2}
\end{figure}

\begin{figure}[htp]
\vspace{1cm}
\centerline{\psfig{file=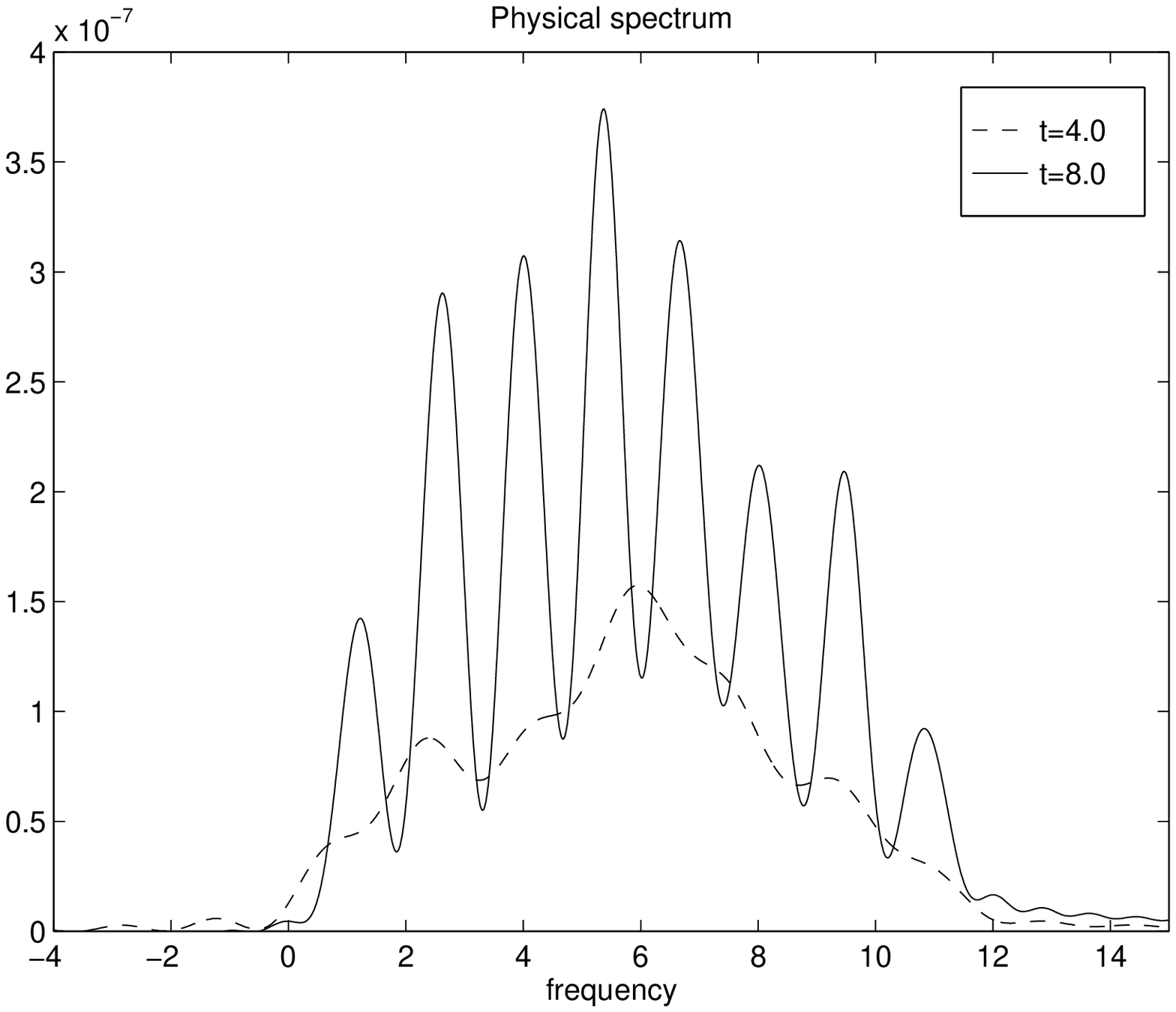,width=8.5cm,bbllx=1cm,bblly=1cm,bburx=20cm,bbury=23cm,clip=}}
\vspace{-2.2cm}
\caption{
The time dependent physical spectrum calculated using QRT at the times $t=4.0$ and $t=8.0$.
The parameters the same as in Fig. \protect\ref{stat8peak}.}
\label{physspec4and8}
\end{figure}

\begin{figure}[htp]
\vspace{-2cm}
\centerline{\psfig{file=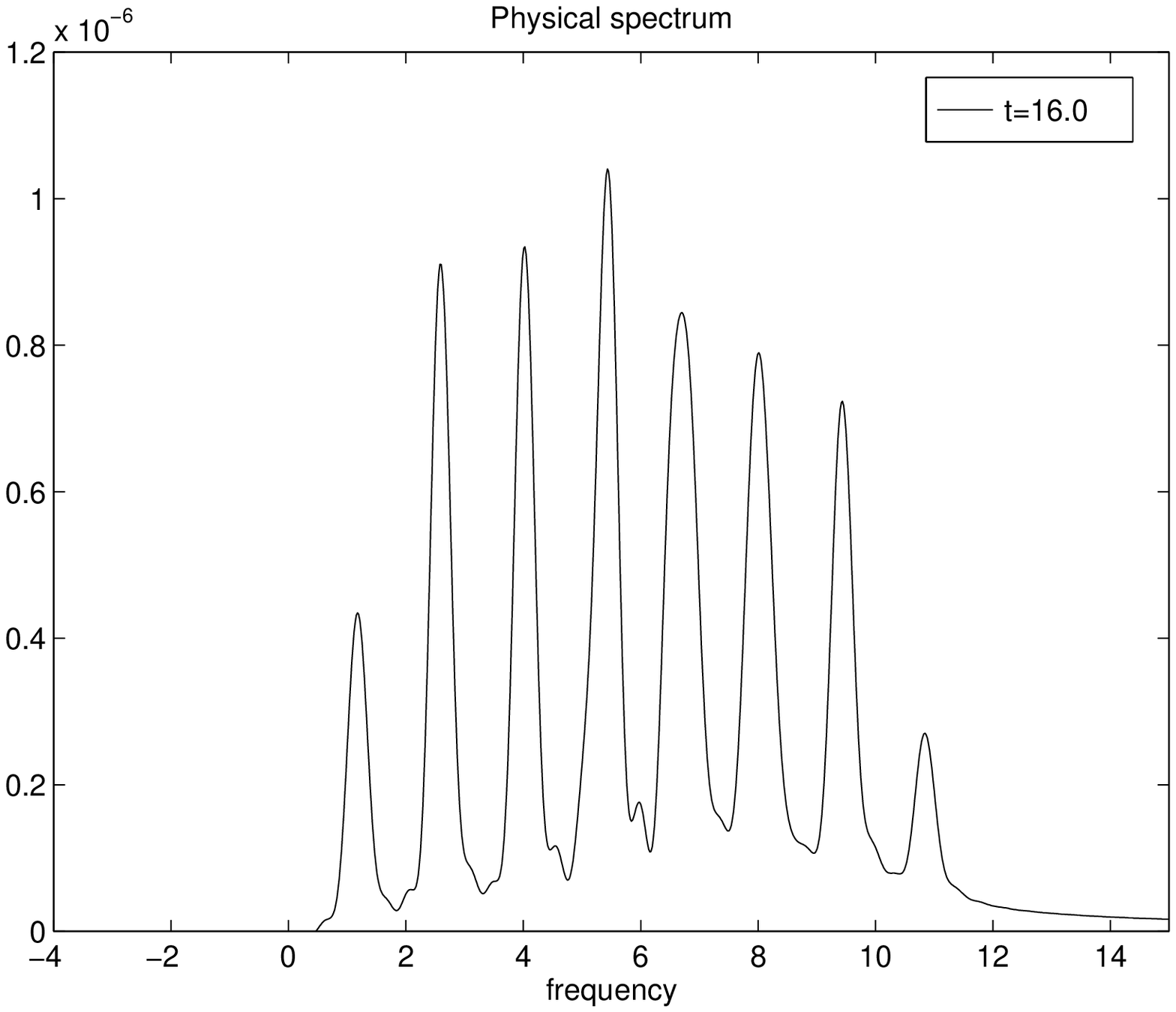,width=8.5cm,bbllx=1cm,bblly=1cm,bburx=20cm,bbury=23cm,clip=}}
\vspace{-2.2cm}
\caption{
The time dependent physical spectrum calculated using QRT at time $t=16.0$.
The parameters the same as in Fig. \protect\ref{stat8peak}.}
\label{physspec16}
\end{figure}

The detection operator used is the same as in the stationary case (\ref{operator1}). Now the expectation values
$\langle\hat{O}^{\dag}(t_1)\hat{O}(t_2)\rangle$ are calculated using the algorithm for a nonstationary
spectrum.
Figures \ref{physspec1and2},\ref{physspec4and8} and \ref{physspec16} show time dependent physical
spectra calculated by the method described above.
The parameters are the same as in Fig. \ref{stat8peak}, so we expect the steady state spectrum to show eight peaks.
The initial state of the system is $\hat\varrho (0)=|2\rangle\langle 2|$.
For the time dependent spectrum also the initial state has some influence. Spectra with different initial
states are different for short times.

From Fig \ref{physspec1and2} figure it is seen that when time is small the spectrum has two broad peaks. At $t=4.0$
a little more structure begins to appear, and at $t=8.0$, eight peaks can be seen. When time increases
the spectrum approaches the steady state spectrum, cf. Fig. \ref{physspec16}. The broad spectrum features
at small times derive from the uncertainty
principle, which does not allow us to see too detailed structures of the spectrum, or more precisely,
the spectrum does not display any detailed structure at small time values.

\subsubsection{The theory of cascaded open system}

\begin{figure}[htp]
\vspace{-1cm}
\centerline{\psfig{file=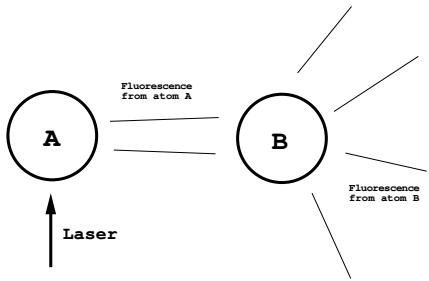,width=8.5cm,bbllx=1cm,bblly=1cm,bburx=20cm,bbury=23cm,angle=-90,clip=}}
\vspace{-1.7cm}
\caption{
The schematic layout of cascaded open systems. Fluorescence radiation from atom A goes to atom B, but
the fluorescence radiation from B does not go back to the system A}
\label{cascadefig1}
\end{figure}

Next we show how the theory of cascaded open systems \cite{gardiner,carmichael}
can be used to calculate the time dependent
spectrum. By this we mean a system (B) which is driven by radiation coming from another
quantum system (A). Radiation from the system A, which may be for example fluorescence light,
is mediated through a reservoir to a system B. The coupling is
uni-directional, the radiation from system B does not go to the system A. The system configuration is shown in
Fig. \ref{cascadefig1}. In the following, we derive a master equation for such a system, following the
presentation in the paper by Carmichael \cite{carmichael}.

The initial Hamiltonian is the following

\begin{equation}
\label{hamilton}
\hat{H}=\hat{H}_A + \hat{H}_B + \hat{H}_R + \hat{H}_{AR} + \hat{H}_{BR},
\end{equation}
where $\hat{H}_A$ and $\hat{H}_B$ are the Hamiltonians for systems $A$ and $B$ respectively.
$\hat{H}_R$ is the reservoir Hamiltonian which mediates radiation from $A$ to $B$.
$\hat{H}_{AR}$ and $\hat{H}_{BR}$ are coupling Hamiltonians to the reservoir

\begin{eqnarray}
\label{couplhams}
\hat{H}_{AR} & = & i\sqrt{\gamma_A}(\hat{a}\hat{E}^{\dag}(0) - {\rm h.c}) \nonumber \\
\hat{H}_{BR} & = & i\sqrt{\gamma_B}(\hat{b}\hat{E}^{\dag}(l) - {\rm h.c}),
\end{eqnarray}
where $\hat{a}$ and $\hat{b}$ are annihilation operator for system A and B respectively.
The coupling to the reservoir is in two different spatial locations for the systems A and B.
We next proceed to eliminate this complication. Using the unitary transformation

\begin{equation}
\label{unitary}
\hat{U}_A(\tau)=\exp[i (\hat{H}_A + \hat{H}_R + \hat{H}_{AR})\tau],
\end{equation}
we obtain an equation for the transformed density matrix $\hat{\varrho}'$ with Hamiltonians which are
now in the standard form

\begin{equation}
\label{newham}
\hat{H}'= \hat{H}_S + \hat{H}_R + \hat{H}_{SR},
\end{equation}
where

\begin{eqnarray}
\label{newhams}
\hat{H}_S & = & \hat{H}_A + \hat{H}_B + \frac{i}{2}\sqrt{\gamma_A\gamma_B}(\hat{a}^{\dag}\hat{b} - 
{\rm h.c}) \nonumber \\
\hat{H}_{SR} & = & i(\sqrt{\gamma_A}\hat{a} + \sqrt{\gamma_B}\hat{b})(\hat{E}^{\dag}(0) - {\rm h.c})
\end{eqnarray}
The master equation can now be derived in the conventional manner by adiabatic elimination of the reservoir.
With these Hamiltonians the derivation gives

\begin{equation}
\label{master2}
\frac{\partial\hat\varrho '}{\partial t}=-i[\hat{H}_S,\hat{\varrho}'] +
\hat{C}\hat{\varrho}'\hat{C}^{\dag} - \frac{1}{2}\hat{C}^{\dag}\hat{C}\hat{\varrho}' - 
\frac{1}{2}\hat{\varrho}'\hat{C}^{\dag}\hat{C},
\end{equation}
where the Lindblad operator $\hat{C}$ is

\begin{equation}
\label{decayoper}
\hat{C}=\sqrt{\gamma_A}\hat{a} + \sqrt{\gamma_B}\hat{b}.
\end{equation}

Next we give a short description how the theory described above can be used for spectrum measurements. As our detector
system we choose a
two level atom which is in the ground state. When we start to drive the atom using radiation with
very low intensity, the excited state of the atom appears with a small probability. The probability of the excited
state is a function of intensity and the spectrum of the radiation. If the atom has a very narrow linewidth, then
only that part of the radiation which is in resonance can excite it. Because excitation is
linearly proportional to the low intensity, the probability of the excited state is proportional to the the
intensity of the incoming radiation at the atomic resonance frequency.

\section{Calculation of the time dependent spectrum for the three level atom}

\subsection{The master equation for spectrum analyzer}

\begin{figure}[htp]
\vspace{-0.5cm}
\centerline{\psfig{file=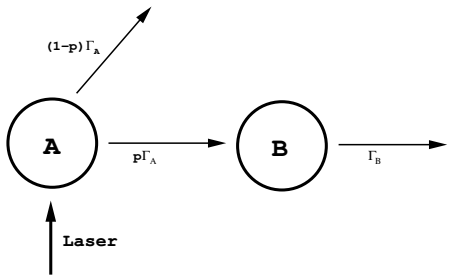,width=8.5cm,bbllx=1cm,bblly=1cm,bburx=20cm,bbury=23cm,angle=-90,clip=}}
\vspace{-2.0cm}
\caption{
This is the version of the cascaded open system we use as a time dependent spectral analyzer.
A small portion $p$ of radiation from atom A goes to atom B. The majority
of the radiation goes into other directions $(1-p)$. The system B decays with decay constant $\Gamma_B$.}
\label{cascadefig2}
\end{figure}

The configuration for a spectrum analyzer is shown in Fig. \ref{cascadefig2}. The system B is a two level atom which is
used to analyze the radiation emitted from the system A. The atom B has a small decay constant $\Gamma_B$ which
corresponds to $\Gamma_F$ in the definition of the physical spectrum (\ref{physspec}). The system A is driven by
a laser. Only a very small fraction of the fluorecent radiation coming from system A is guided to atom B ie. the
constant
$p$ is very small. The majority of the radiation $(1-p)\Gamma_A$ goes into other directions and misses the detector.
Division of the radiation
is needed, because otherwise the incoming radiation would saturate the atom B and excitation would not be
linearly proportional to the intensity any more. Figure \ref{64figure} shows the probabilities of the excited
state of an assembly of analyzer atoms as
functions of time for the system described in detail later in this paper.
Different curves correspond to different values of the resonance frequency.
The spectrum is obtained by plotting the excitation probability at any given time with $\omega_B$ as a parameter.
This method does not need to evaluate multitime averages as in all earlier definitions. The spectrum is obtained
from one time averages of another quantum system, which is coupled to the system we are interested in.
This method of spectrum calculation is quite close to the method presented in the book by 
M.Sargent III et.al. \cite{lamb}.

The system studied was described in section II. The time dependent spectrum of this three level atom has been
calculated using a two level atom as a spectrum analyzer. The master equation for the three level atom and an
analyzer atom according to the theory of cascaded open systems is the following.

\begin{eqnarray}
\label{master3}
\lefteqn{\frac{\partial\hat{\varrho}}{\partial t} = -i[\hat{H}_A + \hat{H}_B + \hat{H}_C,\hat{\varrho}] +}\nonumber\\
& & \hspace{1cm}\sum\limits_{i=1}^4(\hat{L}_i\hat{\varrho}\hat{L}_i^{\dag} - \frac{1}{2}\hat{L}_i^{\dag}\hat{L}_i\hat{\varrho}
-\frac{1}{2}\hat{\varrho}\hat{L}_i^{\dag}\hat{L}_i),
\end{eqnarray}
where 

\begin{eqnarray}
\label{hamiltonabc}
\hat{H}_A & = & \sum\limits_{i=1}^3 E_i|i\rangle\langle i| + (|1\rangle\langle 2|e^{i(\omega_2-\omega_1)t}
 + {\rm h.c}) + \nonumber\\
& & \hspace{2cm}(|2\rangle\langle 3|e^{i(\omega_3-\omega_2)t} + {\rm h.c})  \nonumber \\
\hat{H}_B & = & \frac{1}{2}\omega_B\sigma_z \\
\hat{H}_C & = & \frac{i}{2}\sqrt{\gamma_1p\Gamma_B}(|2\rangle\langle 1|\hat{\sigma}_{-} - {\rm h.c}) + \nonumber\\
\nonumber\\ & & \hspace{1.5cm}\frac{i}{2}\sqrt{\gamma_2p\Gamma_B}(|3\rangle\langle 1|\hat{\sigma}_{-} - {\rm h.c}) \nonumber
\end{eqnarray}
and the Lindblad operators are

\begin{eqnarray}
\label{lindblads}
  \hat{L}_1 & = & \sqrt{\Gamma_{A1}(1-p)}\ |1\rangle\langle 2| \nonumber \\
  \hat{L}_2 & = & \sqrt{\Gamma_{A2}(1-p)}\ |1\rangle\langle 3|  \\
  \hat{L}_3 & = & \sqrt{\Gamma_{A1}p} |1\rangle\langle 2| + \sqrt{0.5\Gamma_B}\ \hat{\sigma}_-^B 
\nonumber \\
  \hat{L}_4 & = & \sqrt{\Gamma_{A2}p} |1\rangle\langle 3| + \sqrt{0.5\Gamma_B}\ \hat{\sigma}_-^B.
 \nonumber
\end{eqnarray}

In addition to the atomic Hamiltonians, there is an additional term in the Hamiltonian arising from
the derivation which gives a small energy shift. The first
two decay operators describe decays of the three level atom which take account of the majority of the radiation. 
They correspond to the $(1-p)\Gamma_A$ term in Fig. \ref{cascadefig2}. The last two terms
describe that part of the radiation which goes to the analyzer atom, the $p\Gamma_A$ term in Fig. \ref{cascadefig2}.
They also include spontaneous emission from atom B.

It can be seen that that if $p=0$ the atoms are not coupled. Then, if the initial state is factorizable
then the two atoms evolve separately. Atom A has two decay terms, which describe the transitions
$|1\rangle\langle 2|$ and $|1\rangle\langle 3|$. The Analyzer atom has one decay term with the decay
constant $\Gamma_B$.
This separation is natural because if $p=0$ there is no radiation coming from atom A to atom B.
If $p$ is nonzero it is possible to trace atom B out of the equations and we get the well known master 
equation for atom A.
This means that system B does not affect the time evolution of system A at all.

\subsection{Simulations studied}

\begin{figure}[htp]
\vspace{-0.3cm}
\centerline{\psfig{file=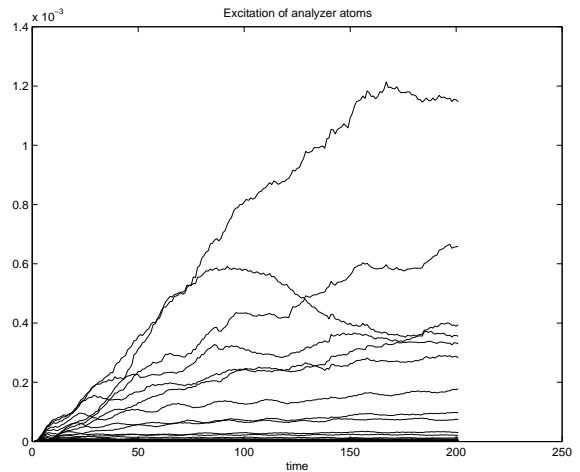,width=8.5cm,bbllx=1cm,bblly=1cm,bburx=20cm,bbury=23cm,clip=}}
\vspace{-2.0cm}
\caption{
Excitation of all 128 analyzer two-level atoms as functions on time. For most of the atoms, the excitations is
so small that it cannot be seen. The three level system parameters are the same as in Fig. \protect\ref{threelevel}.
The laser amplitudes are $\Omega_1=\Omega_2=2.0$, the
parameter $p=0.005$, and the decay constant of the analyzer atoms $\Gamma_B=0.001$. Each curve in this figure
is obtained from an ensemble of identical atoms with 300 members.}
\label{64figure}
\end{figure}

\begin{figure}[htp]
\vspace{-2cm}
\centerline{\psfig{file=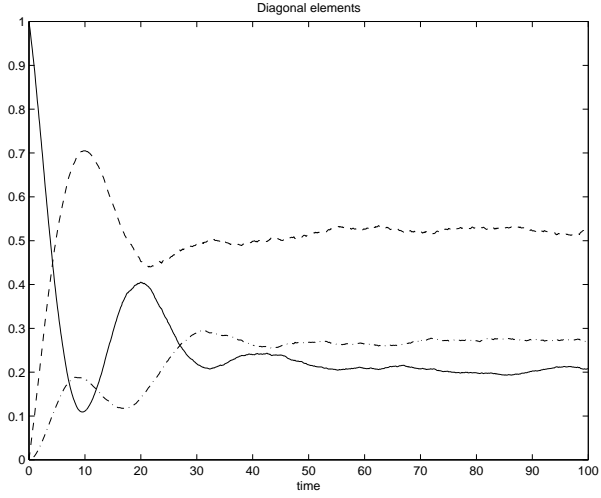,width=8.5cm,bbllx=2cm,bblly=2cm,bburx=20cm,bbury=23cm,clip=}}
\vspace{-2cm}
\caption{
Time evolution of diagonal elements. The energy levels and decay
constants are the same as in Fig. \protect\ref{threelevel}. The laser amplitudes are $\Omega_1=\Omega_2=2.0$.
Dashed line: $\varrho_{11}$, solid line: $\varrho_{22}$, dashdot line: $\varrho_{33}$.}
\label{diagonal}
\end{figure}

\begin{figure}[htp]
\vspace{0cm}
\centerline{\psfig{file=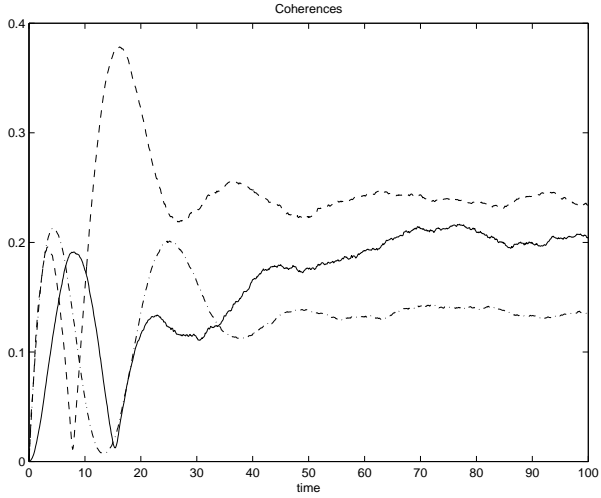,width=8.5cm,bbllx=2cm,bblly=2cm,bburx=20cm,bbury=23cm,clip=}}
\vspace{-2cm}
\caption{
Time evolution of absolute values of off-diagonal elements. The parameters are the same as
in Fig. \protect\ref{diagonal}.
Dashed line: $|\varrho_{12}|$, solid line: $|\varrho_{13}|$, dashdot line: $|\varrho_{23}|$.}
\label{offdiagonal}
\end{figure}

The first case to be studied is our three level system with weak lasers. Figure \ref{diagonal} shows the time
evolution of the diagonal
elements of the density matrix. The initial state is $\hat{\varrho}(0)=|2\rangle\langle 2|$. At first the
population of level
two decreases and starts to Rabi oscillate, at large times it approaches a constant and reaches its steady
state. The populations of levels one and three increase in the beginning and for large times reach steady state.
The behaviour of the off-diagonal elements is quite similar (Fig. \ref{offdiagonal}), first there are
oscillations which smoothen out at large times.
In Fig. \ref{64figure} the population of the excited level is shown for all 128 analyzer atoms used in our simulations.
The different curves belong to different analyzer atoms with different frequencies. They are clearly excited
differently
and this means that the atoms can really recognize the spectral structure of the incoming radiation.
The simulations were done using
the Monte-Carlo wavefunction technique \cite{carmichaelbrussel,mcdprl,mcdjosa,dum}.
The curves are not very smooth so bigger ensembles would be needed.
The numerical calculations were, however, very demanding and 128 different target atoms and the size of their
ensembles were chosen for practical reasons.

\begin{figure}[htp]
\vspace{-2cm}
\centerline{\psfig{file=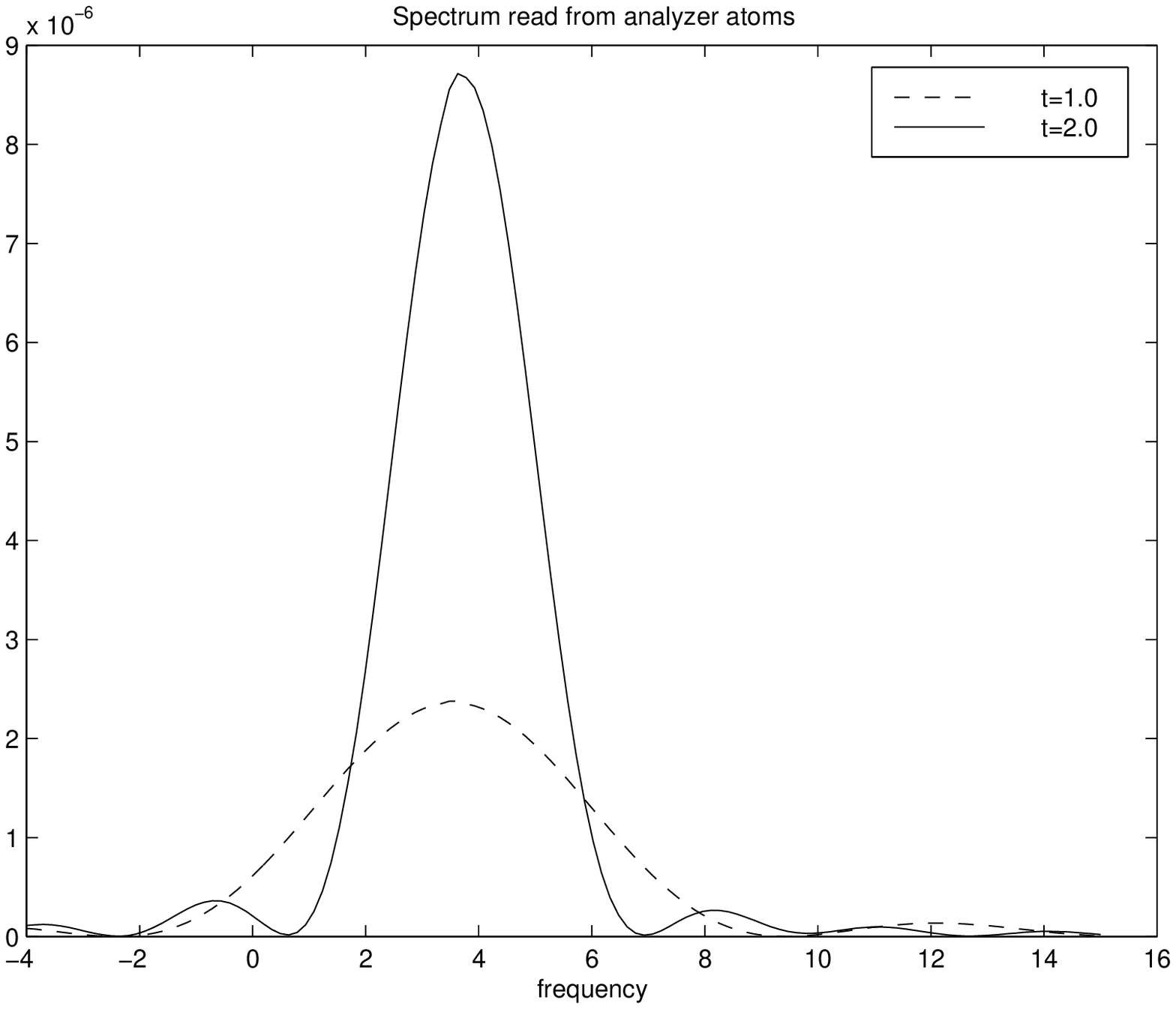,width=8.5cm,bbllx=2cm,bblly=2cm,bburx=20cm,bbury=23cm,clip=}}
\vspace{-2cm}
\caption{
The time dependent spectrum read from the analyzer atoms at times $t=1.0$ and $t=2.0$.
The parameters are the same as in Fig. \protect\ref{stat2peak}.}
\label{anaspecI1and2}
\end{figure}

\begin{figure}[htp]
\vspace{0cm}
\centerline{\psfig{file=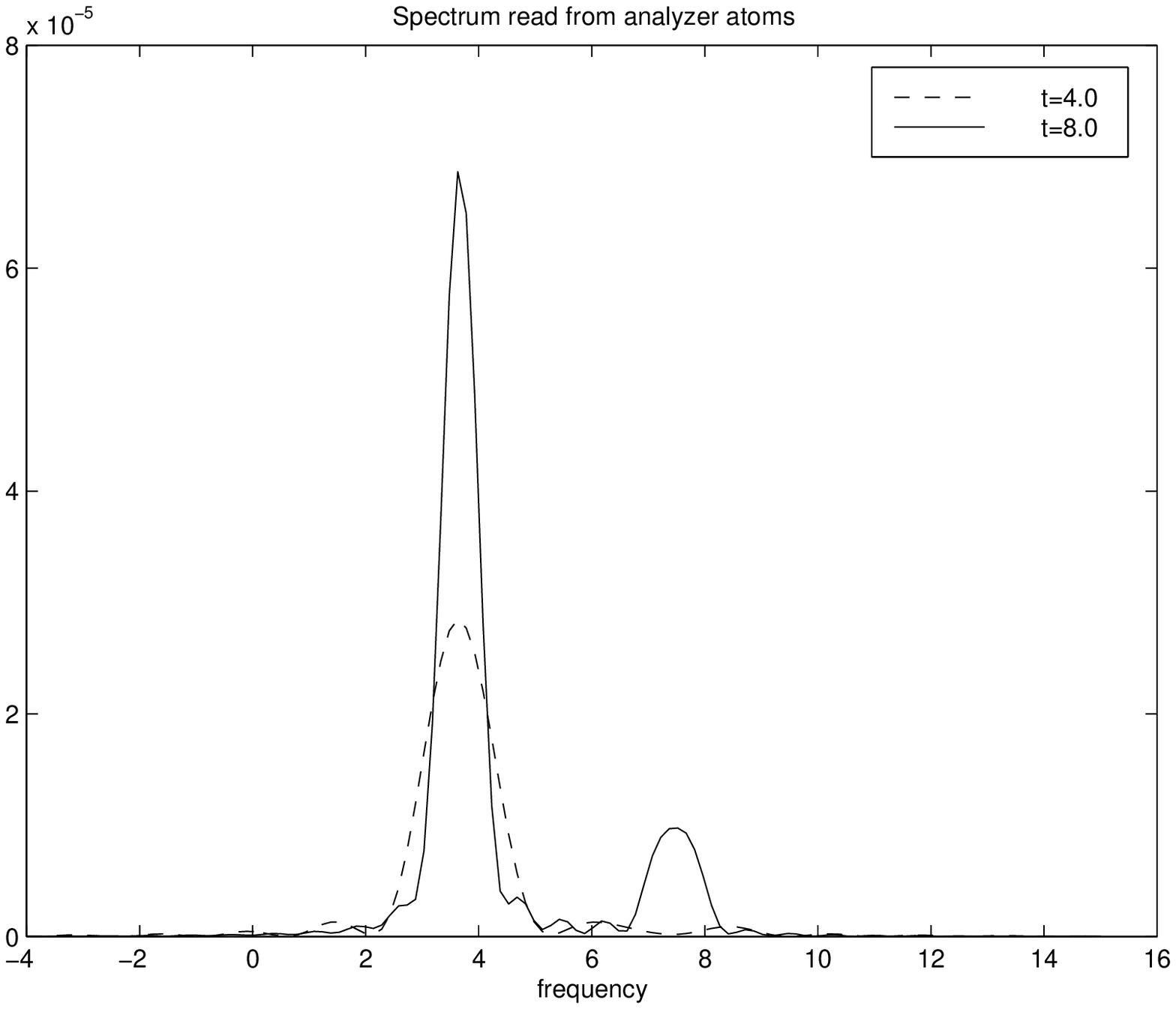,width=8.5cm,bbllx=2cm,bblly=2cm,bburx=20cm,bbury=23cm,clip=}}
\vspace{-2cm}
\caption{
The time dependent spectrum read from the analyzer atoms at times $t=4.0$ and $t=8.0$.
The parameters are the same as in Fig. \protect\ref{stat2peak}.}
\label{anaspecI4and8}
\end{figure}

\begin{figure}[htp]
\vspace{-2cm}
\centerline{\psfig{file=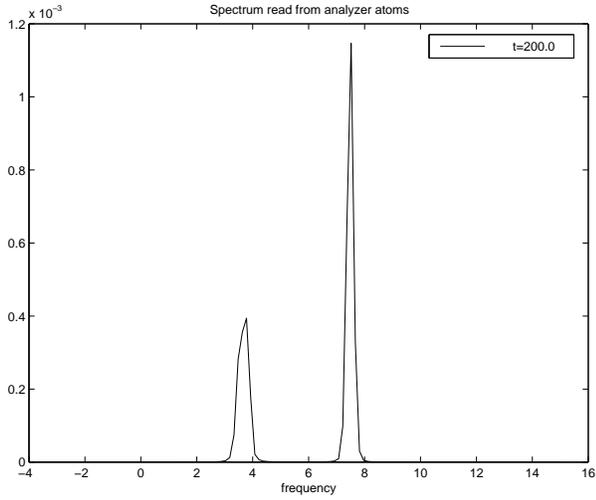,width=8.5cm,bbllx=2cm,bblly=2cm,bburx=20cm,bbury=23cm,clip=}}
\vspace{-2cm}
\caption{
The time dependent spectrum read from the analyzer atoms at time $t=200.0$.
The parameters are the same as in Fig. \protect\ref{stat2peak}.}
\label{anaspecI16and80}
\end{figure}

Spectra read from the analyzer atoms with weak lasers are shown in Fig. \ref{anaspecI1and2},
\ref{anaspecI4and8} and \ref{anaspecI16and80}. The parameters are the same as in Fig \ref{stat2peak}.
The parameter $p$ in the Lindblad operators (\ref{lindblads}) is chosen to be $p=0.005$.
The decay constant for analyzer atoms is $\Gamma_B=0.001$.
The initial state is $\hat{\varrho}(0)=|2\rangle\langle 2|$.

In figure \ref{anaspecI1and2} the spectrum is  broad. At $t=2.0$ there is a broad peak centered around $\omega=4$
and as time increases the peak becomes narrower which can be seen at $t=8.0$ (Fig. \ref{anaspecI4and8}).
The appearance of a peak around $\omega=4$ is understandable,
because initially level two was populated. The spectrum reveals that the atom decays from level two to the
ground level. As time increases, another peak begins to appear at $\omega=8.0$. Its relative intensity
increases and in steady state it is bigger than the $\omega=4.0$ peak. At long times, $t=200$, the spectrum
reaches its steady state Fig. \ref{anaspecI16and80}; cf. Fig \ref{stat2peak}.

\begin{figure}[htp]
\vspace{0cm}
\centerline{\psfig{file=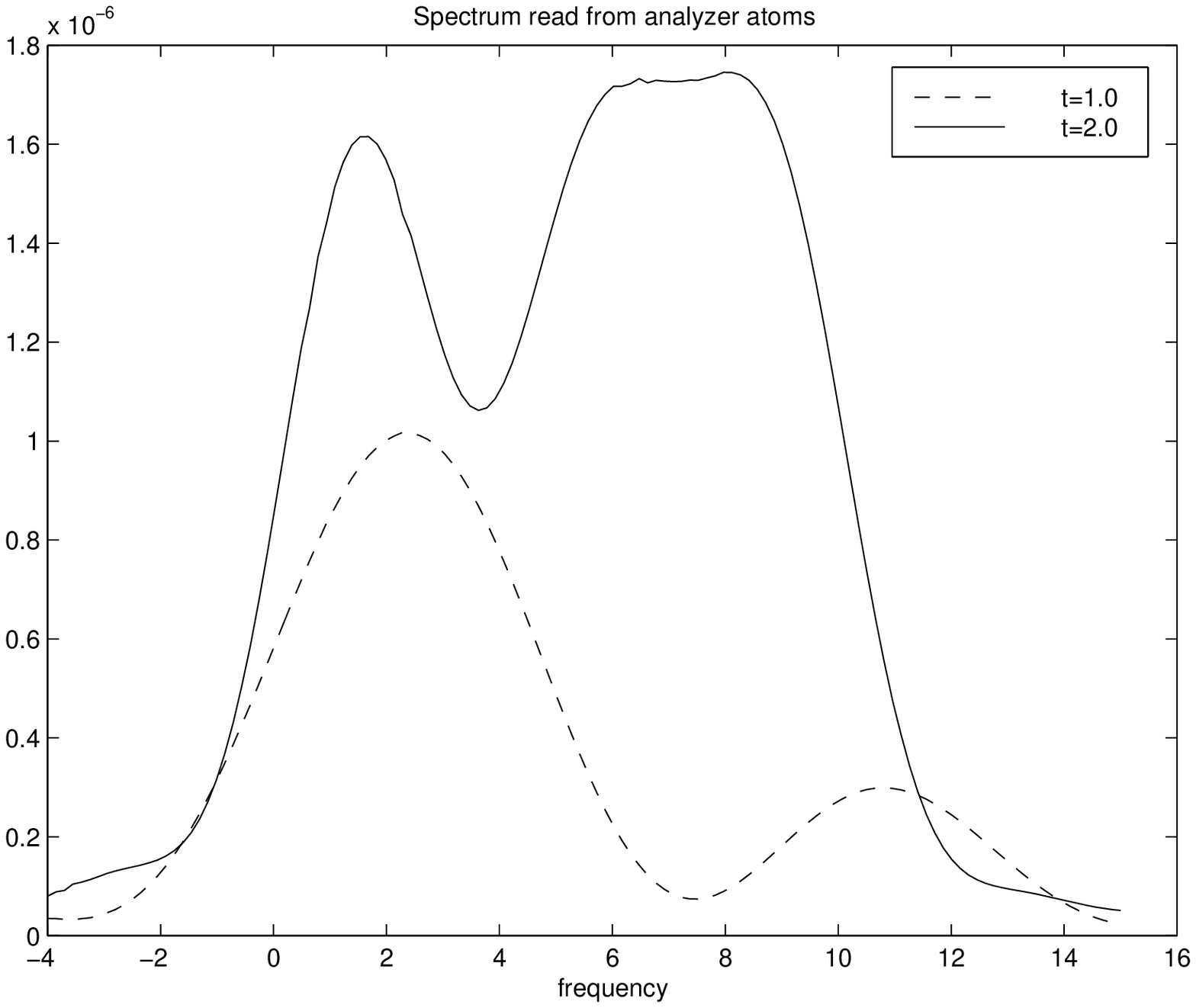,width=8.5cm,bbllx=2cm,bblly=2cm,bburx=20cm,bbury=23cm,clip=}}
\vspace{-2cm}
\caption{
The time dependent spectrum read from the analyzer atoms at times $t=1.0$ and $t=2.0$.
The parameters are the same as in Fig. \protect\ref{stat8peak}. Compare this with Fig. \protect\ref{physspec1and2}
calculated with the same parameters.}
\label{anaspecII1and2}
\end{figure}

\begin{figure}[htp]
\vspace{-2cm}
\centerline{\psfig{file=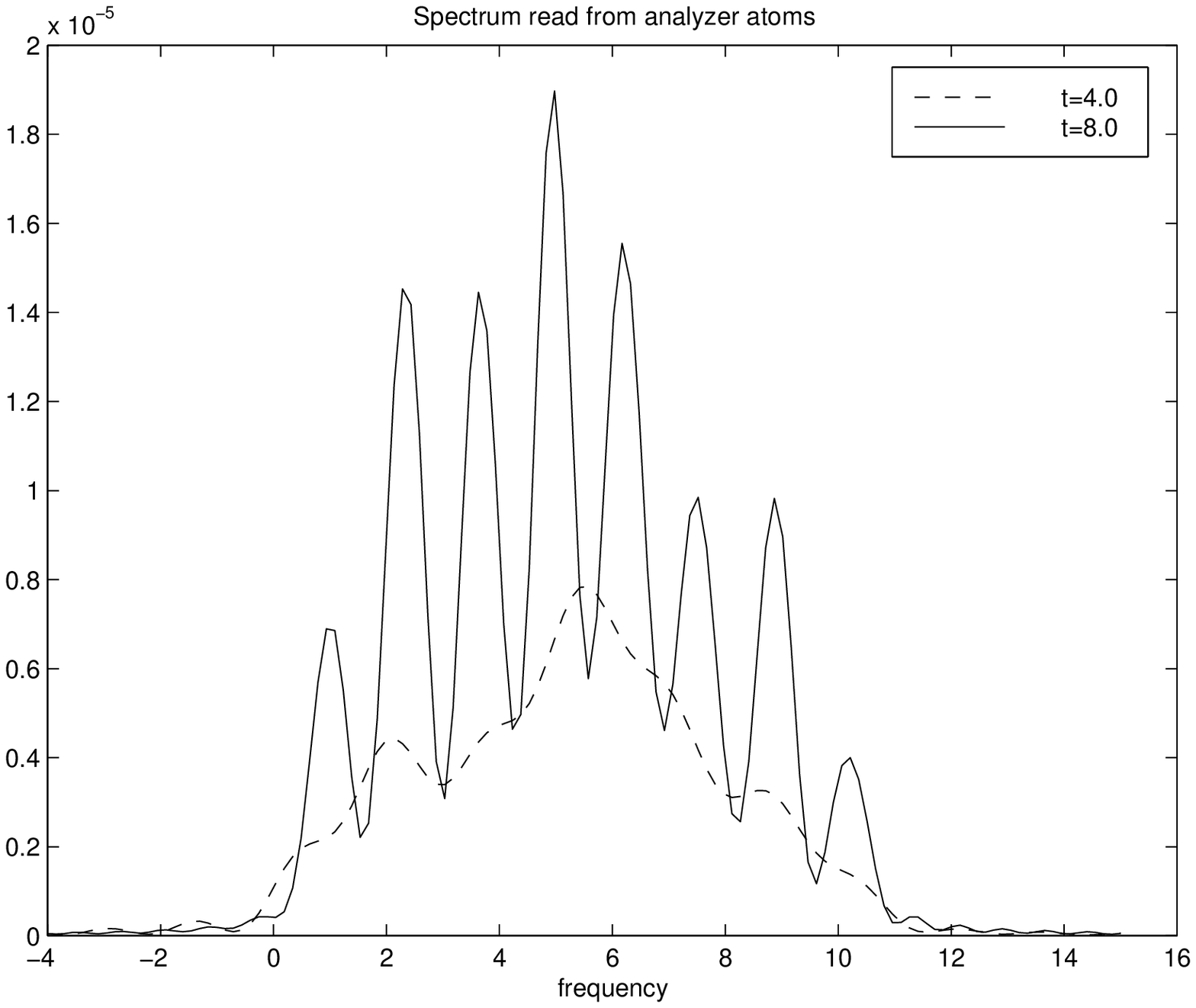,width=8.5cm,bbllx=2cm,bblly=2cm,bburx=20cm,bbury=23cm,clip=}}
\vspace{-2cm}
\caption{
The time dependent spectrum read from the analyzer atoms at times $t=4.0$ and $t=8.0$.
The parameters are the same as in Fig. \protect\ref{stat8peak}. Compare this with Fig. \protect\ref{physspec4and8}
calculated with the same parameters.}
\label{anaspecII4and8}
\end{figure}

\begin{figure}[htp]
\vspace{0cm}
\centerline{\psfig{file=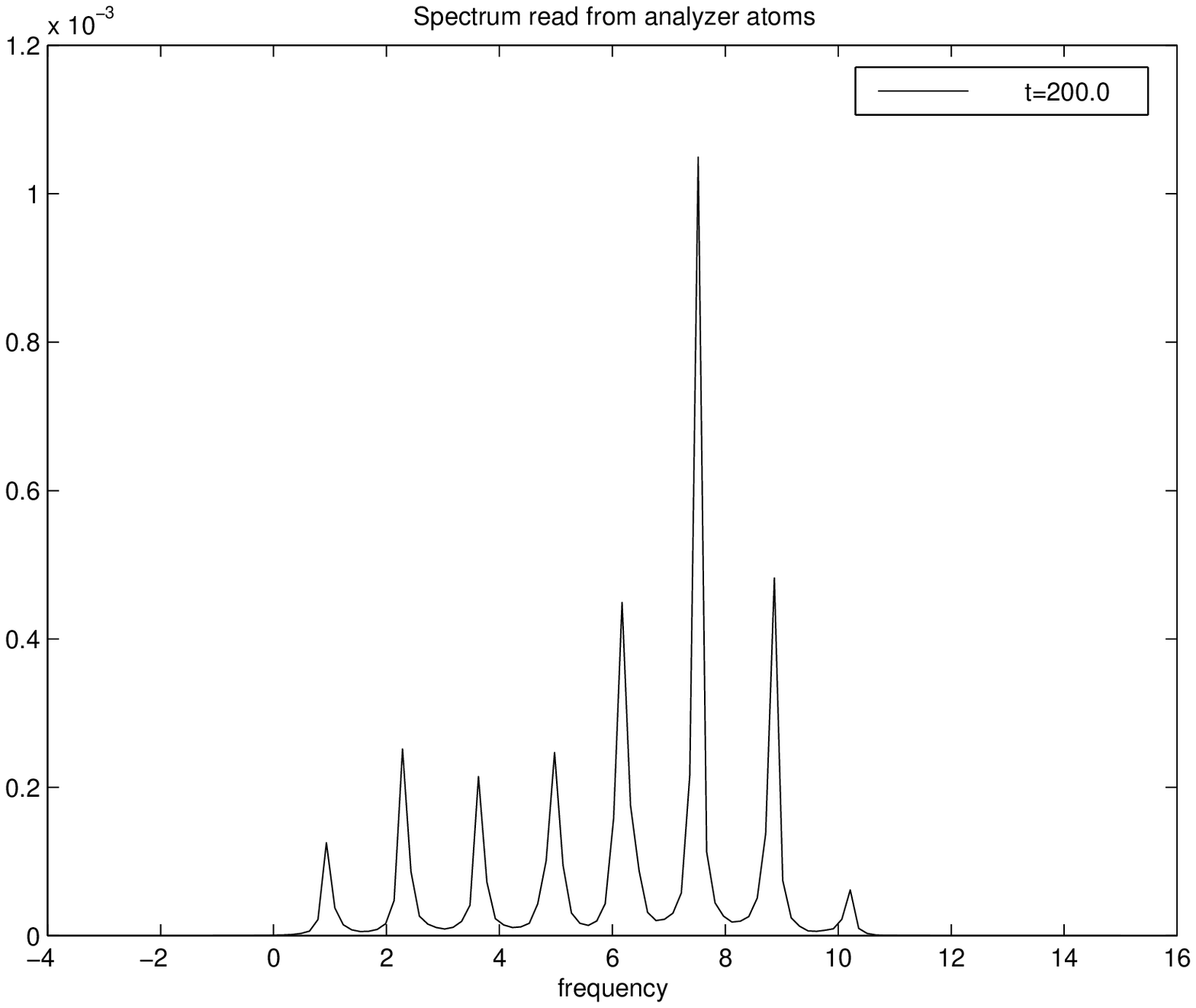,width=8.5cm,bbllx=2cm,bblly=2cm,bburx=20cm,bbury=23cm,clip=}}
\vspace{-2cm}
\caption{
The time dependent spectrum read from the analyzer atoms at time $t=200.0$.
The parameters are the same as in Fig. \protect\ref{stat8peak}. Compare this with Fig. \protect\ref{stat8peak}
calculated with the same parameters.}
\label{anaspecII100}
\end{figure}

The spectra with stronger lasers are shown in figures \ref{anaspecII1and2},\ref{anaspecII4and8} and
\ref{anaspecII100}. At small times, the spectrum is broad and the
more precise structures appears when time increases. In the last figure the time is very large $t=200$
and spectrum has reached its steady state Fig. \ref{stat8peak}.

%Time dependent spectrum when lasers are strong is shown in figures 12,13,14. First
%spectrum is broad and it is centered symmetrically around $\omega =4.0$. This is as expected because initial
%state is $\hat{\varrho}(0)=|2\rangle\langle 2|$. As time increases spectrum becomes narrower. At $t=8.0$
%second peaks appears at $\omega =8.0$ and at stationary limit there are two narrow peaks at $\omega =4.0$
%and $\omega =8.0$ as expected.

%Spectrum calculated form
%analyzer atom is shown in figure (figxx16,figxx17). Spectrum is almost exactly the same as physical spectrum
%calculated from equation (\ref{physspec}) and using QRT.

\newpage

\subsection{Comparison between methods}

The spectrum calculated using analyzer atoms and the definition of the physical spectrum seem to give
very similar results.
However, the calculation techniques are totally different and from the mathematics it is not at all obvious that
these two spectra should agree. For the physical spectrum, two time averages are needed and the spectrum is a
kind of Fourier-transform of them. In this respect the physical spectrum is closer to the Wiener-Khintchine spectrum.
The method with analyzer atoms does not need any multitime-averages. The spectrum is read from a quantum system
(the two state atom) which acts as a measurement device.

It is also interesting to compare how the uncertainty principle restricts spectral measurements. In all our
simulations, analyzer atoms of very small linewidth were used. This means that the atoms respond to the
changes of the incoming radiation very slowly, but if the interaction time is long enough the atoms give
an accurate spectrum. If we
used atoms with a larger linewidth, the atoms would indicate the state of the system more quickly, but because
of their large linewidth they would not give a very accurate spectrum.
In the case of a the physical spectrum, the uncertainty relation is related to the properties of the Fourier transform.

The method of analyzer atoms is based on the simulation of a master equation. Recently many parallel stochastic
algorithms have been presented \cite{carmichael,mcdprl,mcdjosa,dum}. There are also freely available programs which
are developed for simulations of these kinds of master equations \cite{cplusplus}.
It is also possible to parallelize the algorithm by computing the evolution of the different
analyzer atoms on different processors.

\section{CONCLUSION}
\label{conclusion}
We have shown a fundamentally new method to calculate a spectrum. Instead of calculating two-time averages
using the Quantum Regression Theorem
we couple the system to two state atoms which are used as a measurement device. The spectrum is read from
the excitation
probability ie. from one-time averages of the two state atoms. This method allows us to calculate both
the time dependent and stationary spectrum. Comparisons between the spectra calculated show that using this
new method we find similar results as using two-time averages and a Fourier-transform.

Because of the frequency-time uncertainty, the spectral resolution is poor for initial times. However, when the
bare level structure is considerebly modified by the presence of strong laser fields like in Fig. \ref{stat8peak},
it may not be possible to surmise the low intensity structure, Fig. \ref{stat2peak}, which is characteristic of the
system of interest. The the short time spectra, Figs. \ref{physspec1and2} and \ref{anaspecII1and2} may indicate
the number of transitions involved even if their positions (4 and 8 in our case) are very difficult to locate. When
the measurement goes on, more data are collected and the spectral information approaches the steady state as 
closely as we desire for a meaningful measurement.

\section{ACKNOWLEDGEMENTS}

M.Havukainen wants to thank P.Stenius for bringing the program by R.Schack and T.A.Brun to our attention.

\end{document}